\documentclass[letters,usenatbib]{mnras}

\usepackage{newtxtext,newtxmath}
 \usepackage{indentfirst}
\usepackage{graphicx}

\usepackage[T1]{fontenc}

\DeclareRobustCommand{\VAN}[3]{#2}
\let\VANthebibliography\thebibliography
\def\thebibliography{\DeclareRobustCommand{\VAN}[3]{##3}\VANthebibliography}

\usepackage{graphicx}	
\usepackage{amsmath}	

\usepackage{amssymb}	

\title[$M_{WD} - P_{\rm {orb}}$ relation of sdB + WD binaries ]{WD mass and orbital period relation of sdB + He WD binaries}

\author[Zhang et al.]{
Yangyang Zhang,$^{1,2,3}$
Hai-Liang Chen,$^{1,2,4}$\thanks{E-mail: chenhl@ynao.ac.cn}
Heran Xiong,$^{5}$
Xuefei Chen,$^{1,2,4}$
Zhanwen Han$^{1,2,3,4}$
\\
           $^1$Yunnan Observatories, Chinese Academy of Sciences,Kunming 650216, China \\           
           $^2$ Key Laboratory for the Structure and Evolution of Celestial Objects, Chinese Academy of Sciences, Kunming 650216, China\\
            $^3$ University of the Chinese Academy of Science, Beijing 100049, China\\
            $^4$ Center for Astronomical Mega-Science, Chinese Academy of Sciences, 20A Datun Road, Chaoyang District, Beijing, 100012, China\\
            $^5$Research School of Astronomy and Astrophysics, Mount Stromlo Observatory, The Australian National University, ACT 2611, Australia\\
}

\date{Accepted XXX. Received YYY; in original form ZZZ}

\pubyear{2021}

\begin{document}
\label{firstpage}
\pagerange{\pageref{firstpage}--\pageref{lastpage}}
\maketitle

\begin{abstract}

Most subdwarf B (sdB) + Helium white dwarf (He WD) binaries are believed to be formed from a particular channel.
In this channel, the He WDs are produced first from red giants (RGs) with degenerate cores via stable mass transfer and sdB stars are produced from RGs with degenerate cores via common envelope (CE) ejection.
They are important for the studies of CE evolution, binary evolution, and binary population synthesis. 
However, the relation between WD mass and orbital period of sdB + He WD binaries has not been specifically studied.
In this paper, we first use a semi-analytic method to follow their formation and find a WD mass and orbital period relation. Then we use a detailed stellar evolution code to model their formation from main-sequence binaries. We find a similar relation between the WD mass and orbital period, which is in broad agreement with observations.
For most sdB + He WD systems, if the WD mass (orbital period) can be determined, the orbital period (WD mass) can be inferred with this relation and then the inclination angle can be constrained with the binary mass function. In addition, we can also use this relation to constrain the CE ejection efficiency and find that a relative large CE ejection efficiency is favoured. If both the WD and sdB star masses can be determined, the critical mass ratios of dynamically unstable mass transfer for RG binaries can also be constrained.

\end{abstract}




\section{Introduction}

Subdwarf B (sdB) stars are core helium-burning stars with thin hydrogen envelopes, typically $< 0.02\; {\rm M_{\odot}}$. Their effective temperatures are around $20000-40000\;$K, leading to strong emission in the ultraviolet (UV) band \citep{Heber1986, Saffer1994,Heber2009}. They are important for the studies of asteroseismology \citep[e.g.][]{Charpinet1996, Kilkenny1997,Fontaine2003}, the structure of Galaxy \citep{Green1986}, binary evolution and binary population synthesis \citep{Han2002,Han2003,Han2020}. Because of their strong emission in the UV band, they are the main sources of the UV-upturn in the giant elliptical galaxies \citep{Ferguson1991, Brown1995, Brown1997, Dorman1995, Han2007}. It is widely accepted that sdB stars are mainly formed from three channels, i.e. the common envelope (CE) ejection channel, the stable Roche lobe overflow (RLOF) channel, and the double He WDs merger channel \citep{Han2002,Han2003}.

In the stable RLOF and CE ejection channels, the companion stars of the sdB stars can be main-sequence (MS) stars, WDs and neutron stars \citep[e.g.][]{Maxted2001,Napiwotzki1997,Kupfer2015,Wu2018}. \citet{Kupfer2015} have found that most of these WDs in sdB + WD binaries are likely to be He WDs. Most of these sdB + He WD binaries are likely to be produced from a particular channel \citep{Han2002,Han2003}. In this dominant channel, the relatively massive star in a binary consisting of two MS stars evolves faster into a red giant (RG) star with a degenerate He core and has stable mass transfer. At the end of the mass transfer, a He WD + MS binary system is produced. As the system evolves, the secondary star evolves into a RG star and has dynamically unstable mass transfer leading to a CE phase. During the CE phase, the core of the RG and He WD will lose orbital energy and angular momentum, leading to a shrinkage of the orbit. The lost orbital energy is used to ejected the CE. If the system survives from the CE evolution and the core of the RG star is ignited, then the system will evolve into a binary system consisting of a sdB star and He WD. In this channel, the progenitors of sdB stars can have non-degenerate cores. But these systems contribute insignificantly ($\lesssim 10\%$) to the formation of sdB + He WD binaries.

Besides this formation channel, there is another formation channel, which may account for a very small fraction ($\lesssim 1\%$) of sdB + He WD binaries. In this formation channel, the two components of a binary system with a mass ratio close to $1$, are on the red giant branch (RGB) and the system has unstable mass transfer, the system can also evolve into a sdB + He WD binary after CE ejection \citep{Justham2011}.

For a binary system initiating stable mass transfer on the RGB, if the core of the RG star is degenerate, it is well known that there exit a relationship between the core mass of the RG and the orbital period at the end of mass transfer. This relation has been confirmed by many studies of binary millisecond pulsars \citep[e.g.][]{Savonije87,Joss1987,Rappaport1995,Tauris1999} and wide sdB binaries \citep{Chen2013,Vos2019,Vos2020}. This relation stems from the relation between the core masses and the radii of the giant stars with degenerate cores \citep{Refsdal1971, Webbink1983}. Therefore, in the dominant formation channel of sdB + He WD binaries, we will expect a relation between the WD mass and the orbital period at the end of the first mass transfer. But after the CE phase, it is unclear whether there is still a relation between the WD mass and orbital period.

\citet{Han2003} has modelled the formation of sdB stars via different formation channels and studied the distribution of minimum WD mass and orbital period of sdB + He WD binaries (see their fig.~1).  In addition, \citet{Clausen2012} have studied different subpopulation of sdB binaries. They also present the distribution of minimum companion mass and orbital period for different types of sdB binaries (see their fig. 5). In these two studies, they have not specifically studied the sdB + He WD binaries and investigate the relation of WD mass and orbital period for sdB + He WD binaries.

The aim of this paper is to study the relation between the WD mass and orbital period of sdB + He WD binaries with a semi-analytic method and binary evolution. A similar method has been used in \citet{Shen09} and \citet{Nelson2018}, who study novae and precataclysmic variable systems.
The paper is structured as follows. In Section 2, We follow the formation of sdB + He WD binaries with a semi-analytic method and derive the relation between the WD mass and orbital period. In Section 3, we use a detailed binary evolution code to model the formation of sdB + He WD binaries. Then we get the WD mass and orbital period relation. In Section 4, we compare this relation with observational data of sdB + He WD binaries. We summarize our results and have a brief discussion in Section 5.

\section{The WD mass - orbital period relation from a semi-analytic method}
\label{102}

In order to better understand the WD mass and orbital period relation of sdB + He WD binaries, as the first step, we try to derive this relation with a semi-analytic method. 

For a RG star with a degenerate core, it is known that there is a relation between its core mass (${M_\mathrm{c}}$) and radius ($R$), i.e. ${M_\mathrm{c}}$ - $R$ relation \citep{Refsdal1971, Webbink1983, Joss1987}. 
\citet {Rappaport1995} have fitted this relation with the following formula:

\begin{equation}
\label{1}
R =  \frac{{\rm R_{0}}M_{\mathrm {c}}^{4.5}}{1+4M_{\mathrm {c}}^{4}}+0.5 \mathrm{R_{\sun}} \equiv f_{1} (M_{\rm c}),
\end{equation}
where ${\rm R_{0}}$ is a constant.  It equals to 5500 $\rm R_{\sun}$ for Population I (metallicity $Z = 0.02$) and 3300 $\rm R_{\sun}$ for Population II (metallicity $Z = 0.001$). $M_{\rm c}$ is the core mass in units of solar mass, and R is the radius of the RG star in units of solar radius. For convenience, we denote this formula with a function  $f_{1}(M_{\rm c})$. In the following, we will use similar symbols $Y \equiv f(X_{1}, X_{2})$ to show the dependence of one parameter $Y$ on other parameters ($X_{1}$, $X_{2}$). 

The relation between the binary separation ($a$), the mass ratio ($q$) and Roche lobe radius ($R_{\rm L}$) can be given by \citep{Peter1983}:
 \begin{equation}
 \label{2}
R_{\mathrm{L}}/a \approx \frac{0.49q^{2/3}}{0.6q^{2/3}+\mathrm{ln}(1+q^{1/3})} \equiv f_{2} (q).
\end{equation}

During the stable RLOF or at the beginning of a CE phase, as the donor fills its Roche lobe, its radius ($R$) is approximately equal to its Roche lobe radius ($R_{\rm L}$), i.e. $R \approx R_{\rm L}$. If the donor star is a RG with a degenerate core, then we can find a relation between the core mass ($M_{\rm c}$) and binary separation ($a$):
\begin{equation}
\begin{split}
\label{3}
a & = R_{\rm {L}} / f_{2} (q) \approx R / f_{2} (q) = f_{1} (M_{\rm {c}}) / f_{2} (q)\\
                          & \equiv  f_{3} (M_{\rm c}, q).
\end{split}
\end{equation}

For a sdB + WD system, its progenitor system has undergone two mass transfer phases, which start when the mass-donor stars are on the RGB and have degenerate cores. Just before the end of the first mass transfer phase, the mass-donor star still fills its Roche lobe and its mass is close to the WD mass. With Eq.~\ref{3}, we can get the binary separation at the end of the first mass transfer phase,
\begin{equation}
\begin{split}
\label{4}
a_{\rm {f, RLOF}} & = f_{3} (M_{\rm {c, 1}}, q_{1}) \approx  f_{3} (M_{\rm WD}, M_{\rm {1}}/M_{\rm 2}) \\
                          & \approx  f_{3} (M_{\rm WD}, M_{\rm {WD}}/M_{\rm 2}),
\end{split}
\end{equation}
where  $M_{1}$, $M_{\rm c,1}$ are the masses of the progenitor of WD and its core just before the end of the first mass transfer phase, respectively; $M_{2}$ is the secondary (the progenitor of the sdB star) mass just before the end of the first mass transfer phase.

After the first mass transfer phase, the binary separation is relatively large. The effect of gravitational wave radiation, magnetic braking and tides is relatively weak. Therefore, the binary separation will not change significantly, i.e. the binary separation at the end of the first mass transfer phase ($a_{\rm f,RLOF}$) approximately equals to the binary separation at the onset of the CE phase ($a_{\rm i,CE}$). At the beginning of the CE phase, if the progenitor of the sdB star is a low mass star with a degenerate core, its core mass and radius should also follow the $M_{\rm c}$-$R$ relation.
With Eq.~\ref{3}, we can get the binary separation at the onset of the CE phase
\begin{equation}
\label{5}
a_{\rm {f, RLOF}} = a_{\rm i, CE} = f_{3} (M_{\rm {c ,2}}, q_{2}) \approx f_{3} (M_{\rm {sdB}}, M_{\rm {2}}/M_{\rm {WD}}),
\end{equation}
where $M_{\rm {c,2}}$ is the core mass of the progenitor of the sdB star and approximately equals to the mass of the sdB star.

From Eq.~\ref{4} and ~\ref{5}, we can find a relation between the mass of the sdB star and the mass of the WD and progenitor of the sdB star, i.e.
\begin{equation}
\label{6}
M_{\rm {sdB}} \equiv f_{4} (M_{\rm {WD}}, M_{\rm {2}}).
\end{equation}

Regarding the CE process, we adopt the energy budget prescription \citep{Webbink1984}
 \begin{equation}
 \label{7}
E_{\mathrm{bind}} = \alpha_{\mathrm {ce}}(\frac{GM_{\mathrm{WD}}M_{\mathrm{sdB}}}{2a_{\mathrm{f, CE}}} - \frac{GM_{\mathrm{WD}}M_{2}}{2a_{\mathrm{i, CE}}}),
\end{equation}
where $E_{\mathrm {bind}}$ is the binding energy including gravitational energy ($E_{\mathrm {gr}}$) and thermal energy ($E_{\mathrm {th}}$) of the H-rich envelope:
\begin{equation}
 \label{8}
 E_{\mathrm {bind}}  =  -(E_{\mathrm {gr}} + \alpha_{\mathrm {th}} E_{\mathrm {th}}).
\end{equation}
Here $\alpha_{\mathrm {ce}}$ and $\alpha_{\mathrm {th}}$ are the efficiencies of the released orbital energy and thermal energy used to eject the envelope, respectively \citep {Han2002, Han2003}; $G$ is the gravitational constant; the binding energy of the H-rich envelope is a function of $M_{2}$ and the core mass (i.e. the sdB mass) at the beginning of the CE phase for a fixed $\alpha_{\mathrm {th}}$ :
\begin{equation}
 \label{9}
 E_{\mathrm {bind}}  \equiv  f_{5} (M_{2}, M_{\rm {c, 2}}) =  f_{5} (M_{2}, M_{\rm {sdB}}) .
\end{equation}

With a given $\alpha_{\mathrm {ce}}$, from Eq.~\ref{5},~\ref{6},~\ref{7} and~\ref{9}, we can get the binary separation after the CE ejection
\begin{equation}
\begin{split}
\label{10}
a_{\mathrm {f, CE}} & =  \frac{GM_{\mathrm{WD}}M_{\mathrm{sdB}}}{2E_{\mathrm{bind}}/\alpha_{\mathrm {ce}}+GM_{\mathrm {WD}}M_{2}/{a_{\mathrm {i, CE}}}} \\
& \equiv f_{6} (M_{\rm {WD}}, M_{\rm {2}}).
\end{split}
\end{equation}
From Kepler's third law, Eq.~\ref{6} and~\ref{10}, we can get the orbital period after the CE ejection:
\begin{equation}
 \label{11}
 P_{\rm {orb}}  =   \sqrt{\frac{4\pi^{2} a_{\mathrm {f, CE}}^{3}}{G(M_{\mathrm{WD}}+M_{\mathrm{\rm sdB}})} } \equiv f_{7} (M_{\rm {WD}}, M_{2}).
\end{equation}
From this equation, we can find that the final orbital period only depends on $M_{\rm WD}$ and $M_{2}$.

In order to get a quantitative relation between the WD mass and orbital period, we compute some single stellar models to get the binding energies for RG stars with different masses and sdB star masses for different progenitor masses (see Appendix~\ref{App}).

With these results (see Fig.~\ref{B.1}), we can get the possible mass range of He cores in which a He flash will occur after the H-rich envelope is stripped off for a given stellar mass. Similar to \citet{Han2002}, we can find that the mass range of He cores is very small and the He core mass will decrease as the initial star mass increases. Moreover, for a given value of $\alpha_{\rm th}$, we can also compute the binding energies for different stellar models and present the results for some typical examples in Fig.~\ref{A.1}. For any RG with a given mass and core mass, we can interpolate this grid and obtain its binding energy of the envelope. 

\begin{figure}
   \centering
   \includegraphics[width=\columnwidth]{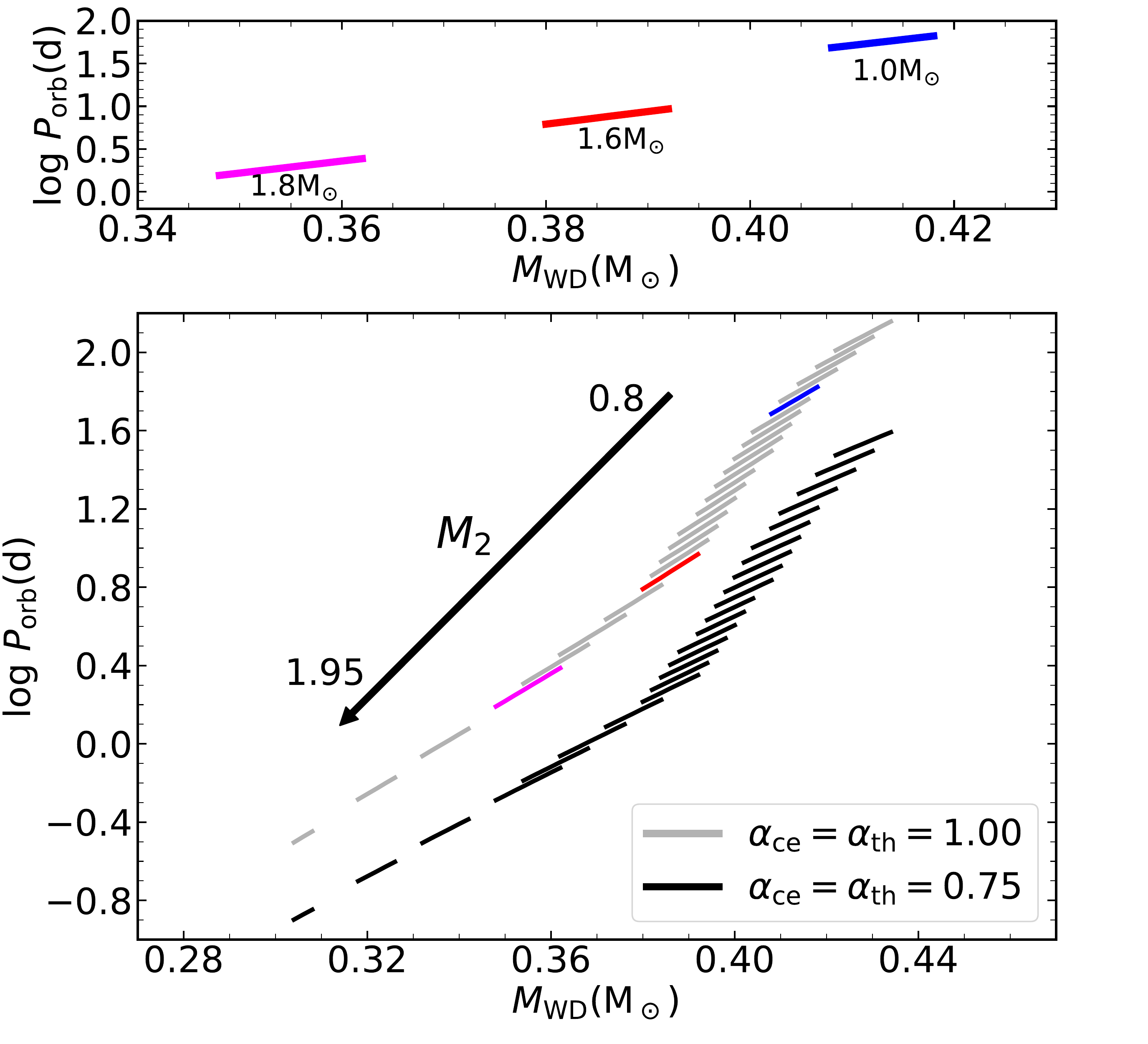}
      \caption{ The WD mass and orbital period relation of sdB + He WD binaries after the CE ejection derived from a semi-analytic method.  In the lower panel, the grey and black lines represent models with $\alpha_{\rm ce}=\alpha_{\rm th} = 1.00$ and $\alpha_{\rm ce}=\alpha_{\rm th} = 0.75$, respectively. Different short lines are for different progenitor masses of sdB stars (i.e. $M_{2}$). 
      The progenitor mass ranges from $0.80\; {\rm M_{\odot}}$ to $1.95\;{\rm M_{\odot}}$ and increases from upper right to lower left. 
      In order to understand the difference of this relation for different $M_{2}$, we choose three relations for $M_{2} = 1.0, 1.60, 1.80\;{\rm M_{\odot}}$, which are highlighted with color lines in the lower panel and shown in the upper panel separately.}
         \label{304}
\end{figure}

With the above calculations, for a given $M_{2}$, we can get the sdB mass from Fig.~\ref{B.1}. With Eq.~\ref{6}, we can get the WD mass. Then we can get the final orbital period from Eq.~\ref{11}. 
In this way, we can obtain the WD mass and orbital period relation for sdB + He WD binaries, which is presented in Fig.~\ref{304}.
From the upper panel of Fig.~\ref{304}, we can find that the range of WD mass is narrow for a given $M_{2}$ and the WD mass is smaller for a larger $M_{2}$. This can be understood. For a given $M_{2}$, the mass range of a sdB star is narrow and the mass of a sdB star is smaller for a larger $M_{2}$ (see Fig.~\ref{B.1}). From Eq.~\ref{6}, we can find that the range of WD mass should also be narrow and the WD mass is smaller for a larger $M_{2}$.
In addition, we can find that the final orbital period is smaller for a larger initial $M_{2}$. For a larger $M_{2}$, it has a more massive H-rich envelope and its core mass at the onset of the CE (i.e. sdB star mass) is smaller. Hence the binding energy of the envelope at the onset of the CE is larger (see Fig.~\ref{A.1}), leading to a smaller orbital period after CE ejection. For larger values of $\alpha_{\rm ce}$ and $\alpha_{\rm th}$, more orbital energy and thermal energy will be used to eject the CE, leading to larger orbital periods after CE ejection. This explains the differences between  models with $\alpha_{\rm ce} = \alpha_{\rm th} = 0.75$ and $\alpha_{\rm ce} = \alpha_{\rm th} = 1.0$.

\section{The WD mass and orbital period relation from detailed binary evolution calculation} 
In order to confirm the existence of this relation, we do realistic binary evolution calculations to model the formation of sdB + He WD binaries from MS binaries with the stellar evolution code \texttt{Modules for Experiments in Stellar Astrophysics} (\texttt{mesa}, version 10398; \citealt{Paxton2011, Paxton2013, Paxton2015, Paxton2018}) in this section.

\subsection{Binary evolution models}

With \textsc{mesa} code, we compute a grid of binary evolution. 
The chemical compositions, mixing length, overshooting, stellar wind, opacity table are the same as we adopted for single stellar evolution models in Appendix~\ref{App}. In the grid, the initial primary star masses are equal to 1.0, 1.26, 1.6, 1.8, 1.95 $\rm M_{\odot}$ at $Z = 0.02$ and 1.0, 1.26, 1.6, 1.85 $\rm M_{\odot}$ at $Z = 0.001$. The secondary masses range from 0.8 to 1.9 $\rm M_{\odot}$ for $Z = 0.02$ and range from 0.8 to 1.8 $\rm M_{\odot}$ for $Z = 0.001$ in steps of $\Delta{M} = 0.1 {\rm M_{\odot}}$. The initial orbital periods range from 3 to 247 days in  logarithmic steps of $0.05$.
 
In order to model their formation as realistic as possible, we consider the angular momentum loss due to gravitational wave radiation\citep{Landau1971}, magnetic braking \citep{Rappaport1983} and mass loss. We adopt the mass transfer scheme of \citet{Kolb1990} to compute the mass transfer rate. In addition, we adopt an isotropic re-emission model for the mass transfer process \citep[see][for a review]{Tauris2006}. 
During the first mass transfer phase, we assume that the mass transfer is completely non-conservative and the lost material leaves the system taking away the specific angular momentum of the accretor. In our calculation, we also assume that the binary will have dynamically unstable mass transfer and enter the CE evolution if the mass transfer rate is larger than $10^{-4}\;\mathrm{M_{\sun} yr^{-1}}$ following \citet{Chen2017}. After the first mass transfer, we take the WD as a point mass and do not consider the evolution of its structure. For the second mass transfer, all the binary systems in our calculation have dynamically unstable mass transfer due to their large
mass ratios (1.7 - 5.6) \citep{Han2002,Chen2008}. We compute the binding energy at the beginning of the second mass transfer with Eq.~\ref{8}. Then we can get the remnant mass of the RG star and binary orbital period after the CE with Eq.~\ref{7},~\ref{10} and~\ref{11} for given values of $\alpha_{\rm ce}$ and $\alpha_{\rm th}$. If the remnant mass is in the range of He core masses for the occurrence of a helium flash (see Fig.~\ref{B.1}), then the core of the RG will evolve into a sdB star. In this way, we can get the WD masses and orbital periods for sdB + He WD systems. We do not follow the subsequent evolution of the binary systems after the CE, which has little influence on the final relation of the WD mass and orbital period.
 
We compute the evolution of all binaries until the onset of second mass transfer or the evolution time is larger than 15 Gyr. Then we find all the binaries which can evolve into sdB + He WD binaries. 

\subsection{Binary evolution results}

\begin{figure}
    \centering
    \includegraphics[width=\columnwidth]{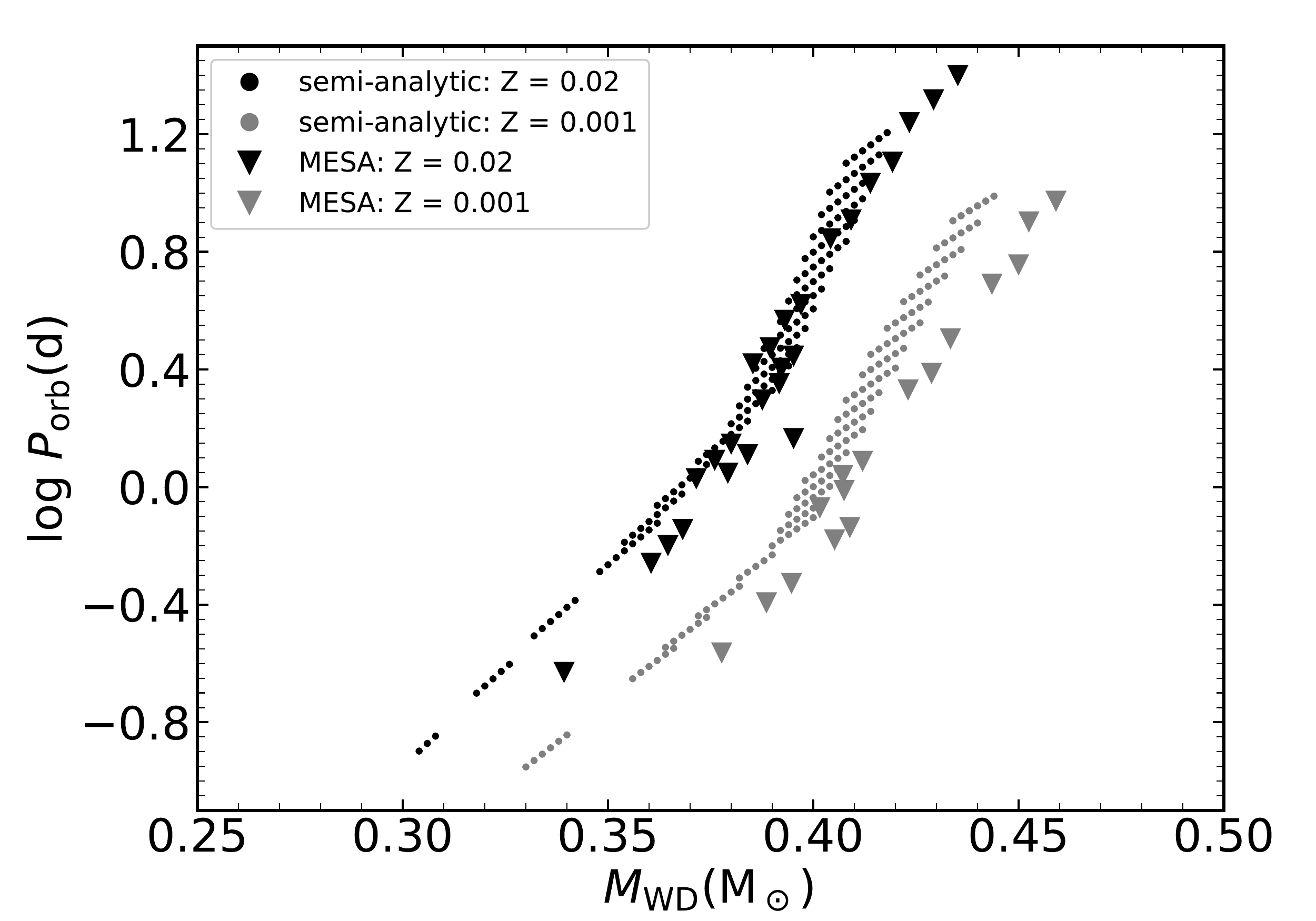}
    \caption{Comparison of WD mass and orbital period relation obtained from the semi-analytic method (dots) and detailed binary evolution calculations (triangles). These black and grey points are for models with $Z = 0.02$ and $Z = 0.001$, respectively. In these models, $\alpha_{\rm ce} = \alpha_{\rm th} = 0.75$ are adopted.}
    \label{fig:mwd_p_bin}
\end{figure}

From the above binary computations, we can get the WD masses and orbital periods for these sdB + He WD binaries, which is shown in Fig.~\ref{fig:mwd_p_bin}. From this plot, we can find that indeed there is a relation between the WD mass and orbital period for sdB + He WDs. For comparison, we also plot the relation obtained with the semi-analytic method. The relations obtained from the semi-analytic method and detailed binary evolution calculations are almost the same. We notice that the maximum orbital periods from these two methods are different. This is mainly because we slightly overestimate the binding energies of the envelopes obtained by interpolation in the semi-analytic method. The minimum orbital periods obtained from \textsc{mesa} calculations is larger, which is mainly due to the maximum initial mass of sdB progenitor stars is smaller compared with models in the semi-analytic method. 
In addition, for the same progenitor mass of sdB stars, the WD masses are larger in \textsc{mesa} calculation compared with that in the semi-analytic method. This can be explained as follows. In the semi-analytic method, the WD mass exactly equals to the core mass of the RG stars, while the WD mass in the \textsc{mesa} calculation includes the H envelope mass. Moreover, at the end of the first mass transfer or at the onset of the CE, the donor radius is smaller than its Roche lobe radius in \textsc{mesa} calculation which is different from that in the semi-analytic method.

\section{Comparison with Observations}
\label{sec:4}
\begin{figure}
    \centering
    \includegraphics[width=\columnwidth]{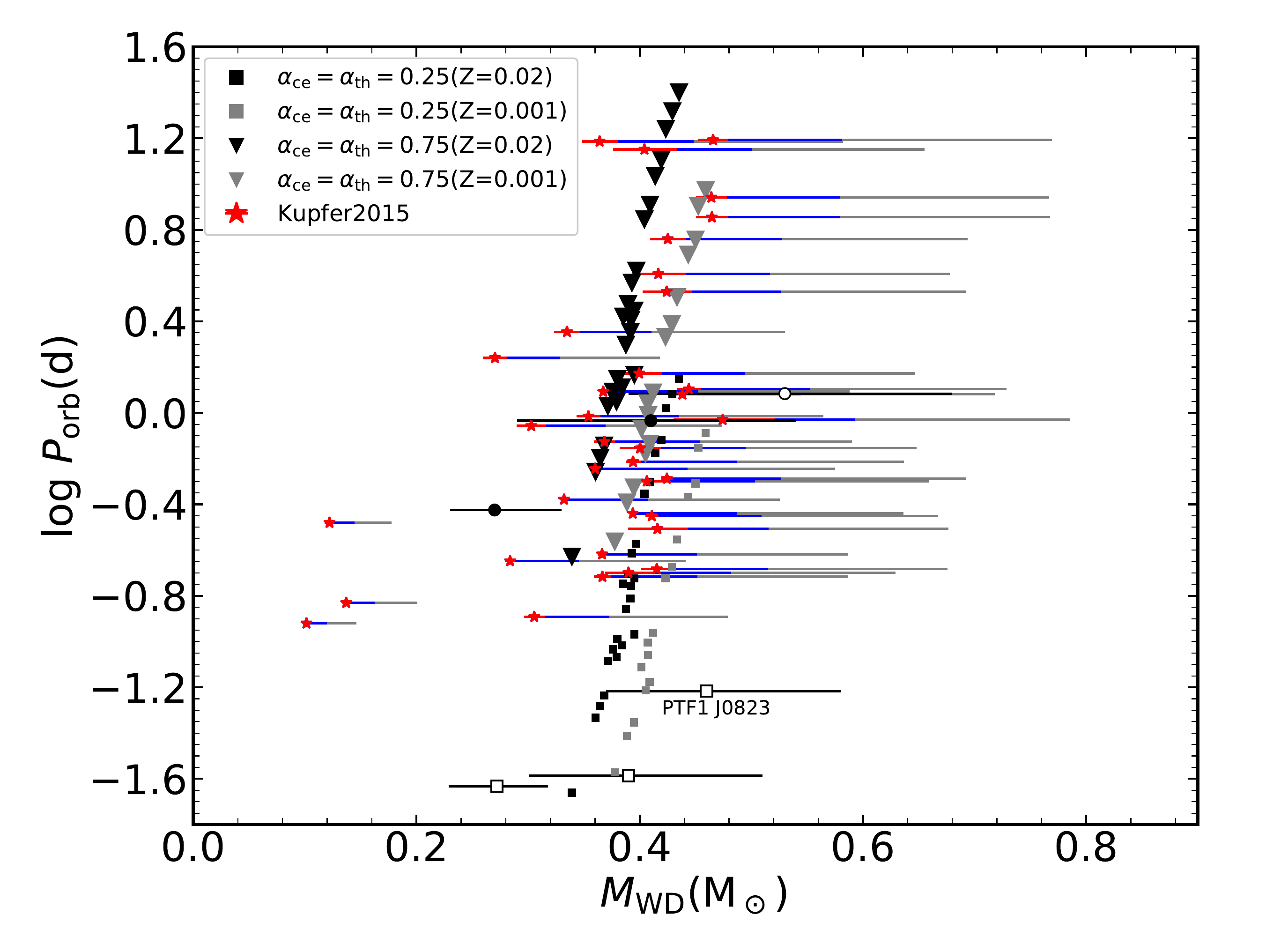}
    \caption{Comparison of WD mass and orbital period relations for sdB + He WD binaries from our calculations with observations. The stars with error bars represent possible He WDs from \citet{Kupfer2015}. The stars indicate the minimum WD masses for the most likely values of orbital periods and RV semi-amplitudes. Given the errors of the orbital periods and RV semi-amplitudes in observations, the red error bars indicate the uncertainties of the minimum WD masses. The gray error bars indicate the uncertainties of orbital inclination angles which follow the theoretical inclination angle distribution ($p(i) = {\rm {sin}}~(i)$). 
    The left end is for an inclination of $90^{\circ}$ and the right end is for an inclination of $46.9^{\circ}$, which corresponds to the $68\%$ (i.e. 1$\sigma$) probability limit.
    The right end of the blue error bars is for an inclination of $60^{\circ}$.
    The black dots, open circles and open squares with error bars represent these systems with well-constrained WD masses \citep{Geier2010a,Kupfer2017,Burdge2020a}. The system with an open circle is likely to have a CO WD. For the systems with open squares, their sdB progenitors likely have non-degenerate cores. The filled squares (triangles) represent the results of \textsc{mesa} models with $\alpha_{\rm {ce}} = \alpha_{\rm {th}} = 0.25$ ($\alpha_{\rm {ce}} = \alpha_{\rm {th}} = 0.75$). The black and gray colors are for models with $Z = 0.02$ and $Z = 0.001$, respectively. 
    }
    \label{fig:com_obs}
\end{figure}

\citet{Kupfer2015} have collected a sample of sdB binaries with identified companion types from the literature (see their table A.3).
In their analysis, they take systems with the following characteristics as sdB binaries with WD companions. 1) These systems show eclipses with no reflection effect and ellipsoidal variation; 2) These systems show no light variations at all; 3) These systems have minimum companion masses larger than the mass for a non-degenerate companion to cause an infrared excess (see Sec. 6 of \citealt{Kupfer2015} for a detailed description). With the above criteria, they found 52 sdB binaries with WD companions. For most systems, due to the uncertainties of inclination angles ($i$), only the minimum WD masses can be determined from the binary mass functions by assuming a critical sdB mass of  $0.47\;{\rm M_{\odot}}$ and $i = 90^{\circ}$.
Excluding these systems with minimum WD masses larger than $0.48\;{\rm M_{\odot}}$ \footnote{The maximum He WD mass is about $0.48\;{\rm M_{\odot}}$ ($0.45\;{\rm M_{\odot}}$) at metallicity $Z = 10^{-4}$ ($Z = 0.02$) \citep[e.g.][]{Han2002,Kilic2007,Parsons2017}. Given the possible metallicity range of this sample, we adopt a maximum He WD mass of $0.48\;{\rm M_{\odot}}$.}, we obtain a sample of 39 sdB + WDs.
Besides the data from \citet{Kupfer2015}, we also collect the data of one sdB + WD system from \citet{Kupfer2017} and another two from \citet{Burdge2020a}, where the inclination angles are estimated and the WD masses are well-constrained.
It is worth noting that some WDs in this sample may be non-pure He WDs (i.e. WDs with CO cores) even their masses are smaller than $0.48\;M_{\odot}$. From observations, it is hard to distinguish this kind of WDs from He WDs.

In Fig.~\ref{fig:com_obs}, we show the comparison of WD mass and orbital period relation of sdB + He WD binaries with these observational data. 
In the plot, these systems without well-constrained WD masses are shown in stars with three types of error bars. The first type (red) error bars indicate the errors of minimum WD masses given the uncertainties of orbital periods and radial velocities (RV) semi-amplitudes in observations. Assuming the inclination angle follows the theoretical inclination angle distribution $p(i) = {\rm {sin}}(i)$ across $0 - 90^{\circ}$, we can get the second type of errors of WD masses ranging from minimum WD masses to 68.3\% (i.e. $i = 46.9^{\circ}$) probability limits (the gray ones). In addition, we obtain the blue error bars by assuming a minimum inclination of $60^{\circ}$.
For these systems with inclination angles estimated and He WD masses well-constrained, we show these observed systems in symbols with black error bars \citep{Geier2010a,Kupfer2017,Burdge2020a}. 

From the plot, we can see that there are some outliers in the sample. We will discuss these system specifically in the following. For the system with an open circle, it is more likely to have a WD mass larger than 0.48 $\rm M_{\odot}$ and could have a CO WD rather than a He WD. Among these observed data, there are three systems with small orbital periods (${\rm log}(P_{\rm orb}/{\rm days}) < -1.2$) (open squares). For the two systems with orbital periods ($\rm {log} (P/\rm {days})$) around -1.6, their sdB masses roughly equal to 0.3 $\rm M_{\odot}$ \citep{Burdge2020a}. Because only these sdB progenitors with non-degenerate cores can ignite the helium when their core masses are around 0.3 $\rm M_{\odot}$ \citep{Han2002,Han2003}. Their progenitor masses of sdB stars are likely to be larger than $2.0\;\rm M_{\odot}$. 
For these progenitors, they have larger binding energy of envelopes leading to smaller orbital periods after CE ejection.  For the system PTF1 J0823, \citet{Kupfer2017} claimed that the progenitor of the sdB star in this system is likely to have a non-degenerate core. For the three systems with minimum WD masses smaller than $0.20\;\rm M_{\odot}$,
it is unclear how the systems formed if the WD masses are confirmed to be less than 0.2 $\rm M_{\odot}$. In principle, stellar objects with such a low mass could be also low-mass main-sequence stars but the fact that no reflection effect in their light curves has been detected excludes this possibility \citep{Maxted2004,Kupfer2015}. 
As we mentioned in the first paragraph of this section, it is possible that some WDs with masses smaller than 0.48 $\rm M_{\odot}$ in the sample are non-pure He  WDs. For these kind of systems, the progenitor of WDs have masses larger than $2.0\;{\rm M_{\odot}}$ and non-degenerate He core ignition on the RGB. After their H envelopes are stripped, they can evolve into WDs with CO cores. With a binary population synthesis method, we estimate that the contribution of low mass ($< 0.48\;{\rm M_{\odot}}$) non-pure He WD + sdB binaries to low mass  ($< 0.48\;{\rm M_{\odot}}$) WD + sdB binaries is less than $5\%$.

From the comparison of our results with observations, we can find that the WD mass and orbital period relation we obtained is in broad agreement with observations. In addition,  we find that a relative large CE ejection efficiency is favoured for the ejection of CE initiating near the tip of RGB.

With the above uncertainties of the formation channel and observational data in mind, a clean sample of sdB + He WDs with well-constrained WD masses and larger orbital periods are needed in order to verify the WD mass and orbital period relation we obtained.

\section{Summary and discussion}

In this paper, we have modelled the formation of sdB + He WD binaries from a particular channel and investigated their properties. In the channel, the He WDs are formed first from RGs with degenerate cores via stable mass transfer and sdB stars are produced from RGs with degenerate cores via CE ejection.
First, with a semi-analytic method, we find that the orbital periods of these systems only depend on the WD masses and the masses of progenitor of sdB stars. In order to confirm this relation, with stellar evolution code \textsc{mesa}, we compute the evolution of a grid of MS binaries to model the formation of sdB + He WD binaries. 
From these calculations, we also find that there exist a similar relation between the WD mass and orbital period of sdB + He WD binaries. Moreover, we compare this relation with observations and find that our results are in broad agreement with observations. In addition, we find that a relatively large value of CE ejection efficiency is favoured.
On the observational side, for most sdB + He WD systems, we can infer the orbital period (WD mass) of sdB + He WD binaries with this relation, if the WD mass (orbital period) can be measured.
Moreover, if both the He WD and sdB star masses can be determined from observations, the progenitor masses of sdB stars can be derived from Eq.~\ref{6}. Then we can get the mass ratios at the onset of CE for these systems. We can use these mass ratios to constrain the critical mass ratios of dynamically unstable mass transfer for RG binaries.

 \section{Acknowledgements}
 
We would like to thank Matthias Kruckow, Jiao Li, Jiangdan Li, Hongwei Ge, Yan Gao for useful comments.  
This work is partially 
supported by the National Natural Science Foundation of China (Grant No. 12090040,12090043,11521303,12073071,11873016,11733008), the Yunnan Province (Grant No. 202001AT070058), 
and the CAS light of West China Program, Youth Innovation Promotion Association of Chinese Academy
of Sciences (Grant no. 2018076).
  
\section{DATA AVAILABILITY}
The data underlying this article will be shared on reasonable request to the corresponding author.




\bibliographystyle{mnras}
\bibliography{zyy_refs}

\appendix

\section{Stellar evolutionary model}
\label{App}

We adopt the state of the art stellar evolutionary code \texttt{Modules for Experiments in Stellar Astrophysics} (\texttt{mesa}, version 10398; \citealt{Paxton2011, Paxton2013, Paxton2015, Paxton2018}) to model the evolution of single stars.
We assume the initial metallicity to be $Z = 0.02$ or $Z = 0.001$. The initial hydrogen abundance is computed as follows $X = 0.76 -3Z$ \citep{Pols1998}.
We set mixing length parameter $\alpha = l/H_{\rm p}$ to be 2.0 (here $l$ is the mixing length and $H_{\rm p}$ is pressure scale height.). Regarding overshooting, we adopt a step-function overshooting model and assume the overshooting region extends a distance of ${0.25 H_{\rm p}}$ \citep{Pols1998, Hurley2000, Han2002}.
We use the prescription from \citet{Reimers1975} with $ \eta = 0.25$ for stellar wind. Furthermore, we consider OPAL type II opacity table \citep{Iglesias1996}.

\begin{figure}
   \centering
   \includegraphics[width=7cm]{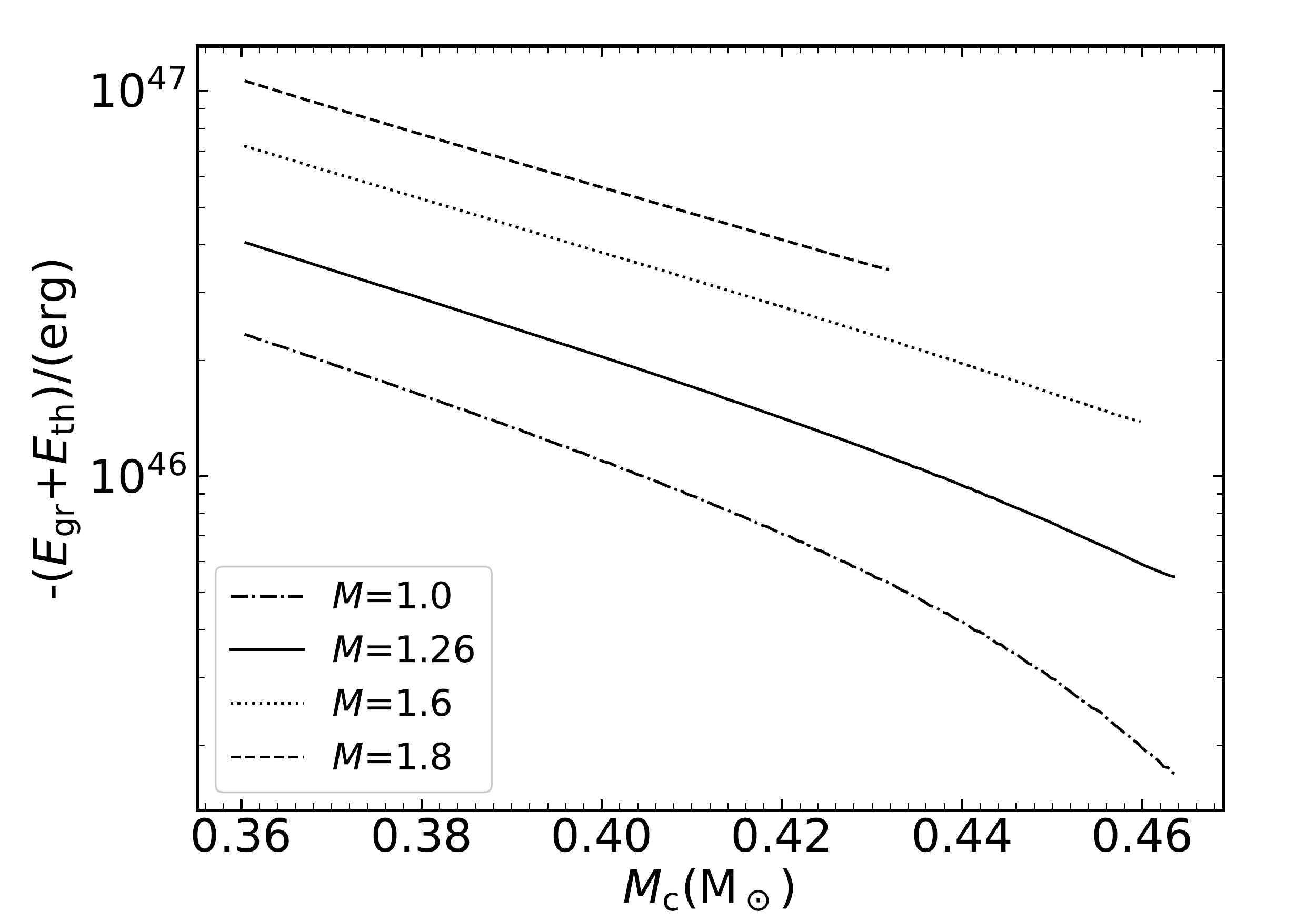}
      \caption{Binding energy as a function of He core mass for four models with typical initial masses: 1.0, 1.26, 1.6, 1.8 $\rm{M_{\sun}}$ at Z = 0.02. In these models, $\alpha_{\rm th} = 1.00$ are adopted. 
           }
         \label{A.1}
\end{figure}

In order to compute the binding energies of the envelopes of RG stars, we compute a grid of stellar evolution tracks. The initial star mass are 1.0, 1.26, 1.6, 1.8 $\rm M_{\sun}$. With Eq.~\ref{8} and $\alpha_{\rm th} = 0.75$ (or $1.00$), we can get the dependence of binding energy on the core mass of the RG. We show some typical models with $\alpha_{\rm th} = 1.00$ in Fig.~\ref{A.1}.

\begin{figure}
   \centering
   \includegraphics[width=7cm]{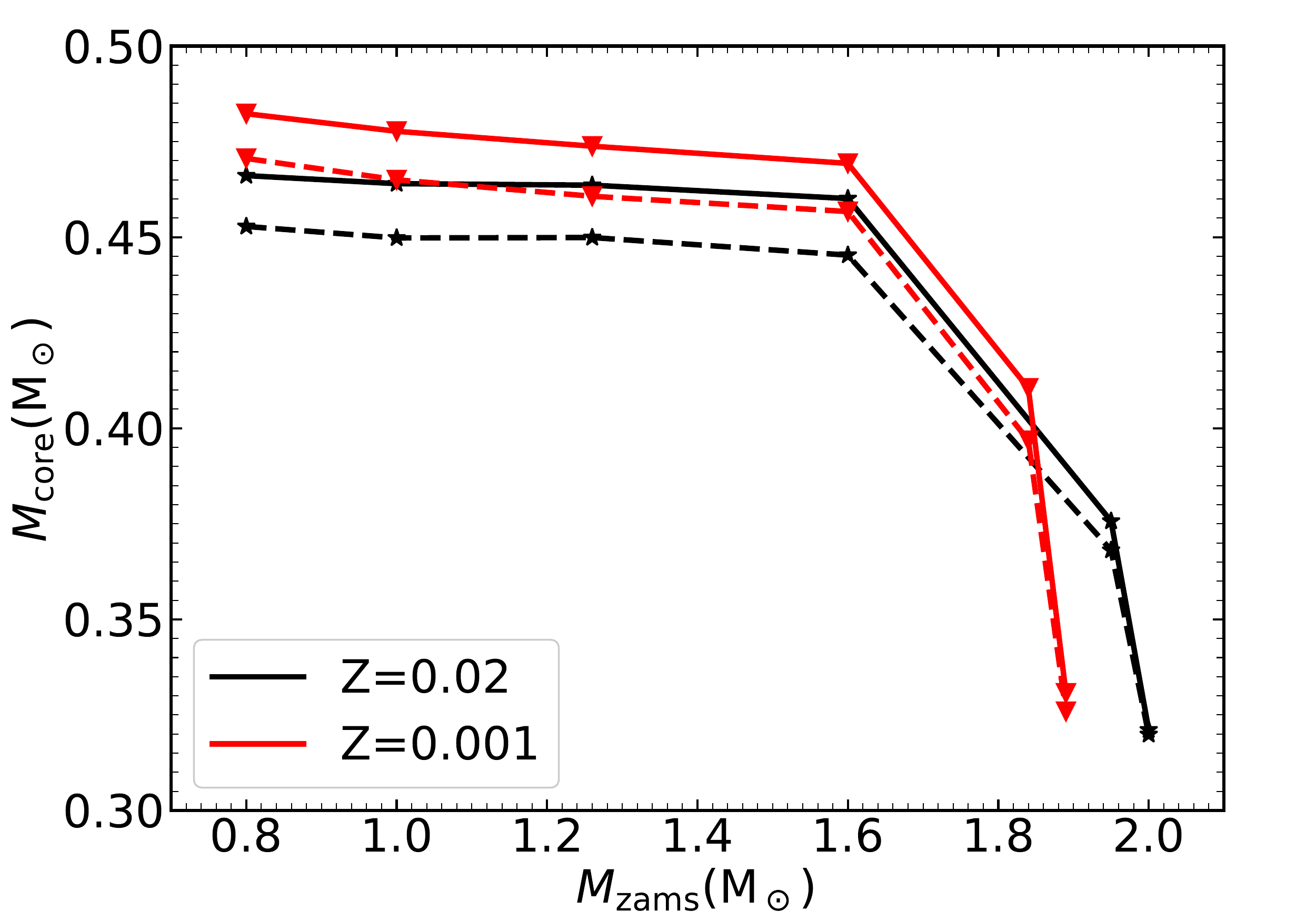}
      \caption{Range of He core masses for the occurrence of a helium flash as a function of the initial mass at $Z = 0.02$ (black color) and $Z = 0.001$ (red color). The dashed line represents the minimum core mass for the occurrence of a helium flash and the solid line shows the core mass at the tip of RGB. Here we only include these stars which have degenerate cores on the RGB.    
           }
   \label{B.1}
\end{figure}

We use the method from \citet{Han2002} to find the range of He core mass for the occurrence of a helium flash. With \texttt{mesa} code, we make a series of models near the tips of RGBs for stars with masses of 0.8, 1.0, 1.26, 1.6, 1.95, 2.0 $\rm M_{\sun}$ at $Z = 0.02$ and 0.8, 1.0, 1.26, 1.6, 1.84, 1.89 $\rm M_{\sun}$ at $Z = 0.001$,respectively. We remove their envelopes with a high mass loss rate until their envelopes collapse.  Then we follow the evolution of the remnant to find whether their central He can be ignited. In this way, we can find the range of core masses for the occurrence of a He flash for different initial masses at $Z = 0.02$ and $Z = 0.001$, respectively, which is presented in Fig.~\ref{B.1}. Our results are consistent with \citet{Han2002}.

\bsp	
\label{lastpage}
\end{document}